\documentclass[aps,eqsecnum]{revtex4}
\usepackage{epsfig}
\usepackage{latexsym}
\usepackage{bm}

\begin{document}
%======================================%
%<<<<<<<<<<<< TITLE PAGE >>>>>>>>>>>>>>%
%======================================%
\thispagestyle{empty}

%
% Osaka University Heading
%
\leftline{\epsfbox{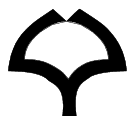}}
\vspace{-10.0mm}
{\baselineskip-4pt
\font\yitp=cmmib10 scaled\magstep2
\font\elevenmib=cmmib10 scaled\magstep1  \skewchar\elevenmib='177
\leftline{\baselineskip20pt
\hspace{12mm} % for revtex
\vbox to0pt
   { {\yitp\hbox{Osaka \hspace{1.5mm} University} }
     {\large\sl\hbox{{Theoretical Astrophysics}} }\vss}}

%
% Preprint numbers
%
{\baselineskip0pt
\rightline{\large\baselineskip14pt\rm\vbox
        to20pt{\hbox{OU-TAP-207}
               \hbox{OCU-PHYS-201}
               \hbox{AP-GR-11}
               \hbox{YITP-03-36}
\vss}}
}
\vskip15mm

\begin{center}

{\Large\bf An Effective Search Method for Gravitational Ringing of Black Holes}

\bigskip

Hiroyuki Nakano$^1$, Hirotaka Takahashi$^{2,3}$, 
Hideyuki Tagoshi$^3$ and Misao Sasaki$^4$

\smallskip
$^1${\em Department of Mathematics and Physics,~Graduate School of 
 Science,~Osaka City University,\\ Osaka 558-8585, Japan
}

\smallskip
$^2${\em Department of Physics,~Graduate School of Science and Technology,
Niigata University,\\ Niigata 950-2181, Japan}\\

\smallskip
$^3${\em Department of Earth and Space Science,~Graduate School of 
 Science,~Osaka University,\\ Toyonaka, 560-0043, Japan
}

\smallskip
$^4${\em Yukawa Institute for Theoretical Physics,
Kyoto University,\\ Kyoto 606-8502, Japan}\\
\smallskip

\medskip

\today

\end{center}

\bigskip

We develop a search method for gravitational ringing of black holes. 
The gravitational ringing is due to complex frequency modes
called the quasi-normal modes that are excited when a black hole 
geometry is perturbed. The detection of it will be
a direct confirmation of the existence of a black hole.
Assuming that the ringdown waves are dominated by the least-damped 
(fundamental) mode with least imaginary part, we consider matched filtering
and develop an optimal method to search for the ringdown waves
that have damped sinusoidal wave forms. 

When we use the matched filtering method, 
the data analysis with a lot of templates required. 
Here we have to ensure a proper match between the filter as a template 
and the real wave. 
It is necessary to keep the detection efficiency as high as possible 
under limited computational costs. 

First, we consider the white noise case 
for which the matched filtering can be studied analytically.
We construct an efficient method for tiling the template space.
Then, using a fitting curve of the TAMA300 DT7 noise spectrum,
we numerically consider the case of colored noise. 
We find our tiling method developed for the white noise case 
is still valid even if the noise is colored.

%%%%%%%%%%%%%%%%%%%%%%%%%%%%%%%%%%%
%%%%%%%%%%%%%%%%%%%%%%%%%%%%%%%%%%%
\section{Introduction}
%%%%%%%%%%%%%%%%%%%%%%%%%%%%%%%%%%%
%%%%%%%%%%%%%%%%%%%%%%%%%%%%%%%%%%%

There are many on-going projects of gravitational wave detection
in the world; LIGO\cite{LIGO}, VIRGO\cite{VIRGO},
GEO-600\cite{GEO}, ACIGA\cite{ACIGA} and TAMA300\cite{TAMA} 
which are ground-based laser interferometers,
and EXPLORER\cite{EXPLORER}, ALLEGRO\cite{ALLEGRO}, NIOBE\cite{NIOBE}, 
NAUTILUS\cite{NAUTILUS} and AURIGA\cite{AURIGA} which are bar detectors. 
Furthermore, there are some future space interferometer projects such 
as LISA\cite{LISA}.

Detection of gravitational waves provides us with not only a direct
experimental test of general relativity but also
a new window to observe our universe.
To use them as a new tool of observation, it is necessary to
know theoretical waveforms. Once we know them,
we may appeal to the matched filtering technique
to extract source's information from gravitational wave signals.
However, because the signals are expected to be
very weak and the amount of data will be enormous
for long-term continuous observations, it is essentially important 
to develop efficient data analysis methods.

For the ground-based and future space-based interferometers, 
the coalescences of compact object binaries 
are the most important sources of gravitational waves. 
The process of coalescence can be divided into three distinct phases. 
During an {\it inspiral} phase, the gravitational radiation 
reaction timescale is much longer than the orbital period. 
The gravitational waves from the inspiral carry the information of 
the masses and spins of the systems and so on. 
After the inspiral, compact object binaries encounter the
dynamical instability and then would merge. 
This phase is called as a {\it merger} phase. 
The gravitational waves from the merger give us 
the information about the highly nonlinear dynamics of relativistic gravity. 
Finally, if a black hole is formed by merger, 
this system can be describe as oscillations of this final 
black hole's quasinormal modes and then 
settles down to a stationary Kerr state. 
The emitted gravitational waves in this {\it ringdown} phase carry 
the information about the mass and spin of the final black hole. 

In this paper, we consider gravitational ringing
of distorted spinning (Kerr) black holes.
The ringdown waves are due to quasi-normal modes of
black holes that are complex frequency wave solutions
of the perturbed Einstein equations with purely outgoing-wave
boundary condition at infinity and ingoing-wave at horizon,
with vanishing incoming-wave amplitude.
A quasi-normal mode is characterized by the central frequency $f_c$,
usually called the (quasi-)normal-mode frequency, 
and the quality factor $Q$ which is inversely proportional to 
the imaginary part of the complex frequency.
The gravitational waves emitted at the last stage of the 
formation of a black hole are also expected to be dominated by 
the quasi-normal modes.

The quasi-normal modes can be obtained by solving the Teukolsky equation 
that governs the perturbation of a Kerr black hole.
Their properties were analyzed extensively by Leaver \cite{Leaver},
and it is known that the least-damped mode belongs to the 
$\ell=m=2$ spin-$2$ spheroidal harmonic modes. 
The dependence of the parameters $\{f_c,\,Q\}$ of the least-damped mode 
on the black hole parameters $\{M,\,J\}$, 
where $M$ is the black hole mass and $J$ is the spin
angular momentum, is briefly reviewed in Sec.~\ref{sec:FQM}. 
Assuming that the ringdown waves are dominated by the
least-damped mode, the black hole parameters $\{M,\,J\}$ can be
uniquely determined by measuring the parameters 
$\{f_c,\,Q\}$ of the gravitational wave signal.

It is noted that the ringdown signal decays exponentially. 
If the signal to noise ratio (SNR) is large, 
not much effort is needed to detect the signal, 
but if the SNR is small, the signal is going to be buried 
inside the noise even more after just one wave length, 
and there is absolutely no hope of seeing many more cycles, 
specifically because the signal drops exponentially with time.
So, we have to treat the loss of the SNR not to be large. 

Methods to search for the ringdown waves were discussed previously
by several authors.
Echeverria~\cite{ech} investigated 
the problem of extracting the black hole parameters
from gravitational wave data in the case when 
the signal-to-noise ratio (SNR) is large. 
Finn \cite{finn} improved this situation by developing 
a maximum likelihood analysis method that can deal
with any SNR.
Flanagan and Hughes then considered the parameter extraction
from the three stages of a binary coalescence,
i.e., from inspiral, merger and ringdown phases,
in their series papers \cite{FH}. 
For the ringdown phase, they discussed the relation between
the energy spectrum of the radiation and the SNR. 
Creighton \cite{crei} reported the results of analyzing 
data of the Caltech 40m by matched filtering,
and emphasized the importance of coincidence event searches
to discriminate spurious events from real events.
But the search was limited to a single ringdown wave template.
In order to treat ringdown waves with unknown parameters, 
we need to prepare a lot of theoretical templates. 
It is necessary to keep the loss of the SNR as small as possible 
under limited computational costs. 
So, we should consider an effective template spacing. 
Recently, Arnaud et al. \cite{arna} have 
discussed a tiling method to cover the 2-dimensional template space 
$\{f_c,\,Q\}$. 
In this paper, we develop a different tiling method which is
much more efficient than that of \cite{arna}
and examine the efficiency with TAMA300 DT7 noise spectrum.

Here we make a comment on combining this search technique 
of ringdown waves with the current search techniques 
for the earlier two stages of the process in the coalescence of 
compact object binaries, i.e., inspiral and merger phases. 
Ultimately, the search method and the data analysis approach should be able 
to handle all the three phases in a unified manner, 
going smoothly from one phase to the next. 
In the present state, 
it may be difficult to obtain the parameters of the final black hole 
from the gravitational waves in the inspiral and merger phases.
However, in the case when a compact star is inspiralng into a
super massive black hole, we will be able to obtain the information 
about the mass of a super massive black hole 
from the gravitational waves in the inspiral phase.
This will improve the detection efficiency of ringdown waves
significantly.

The paper is organized as follows.
In Sec.~\ref{sec:FQM}, 
we briefly review the quasi-normal modes of spinning
black holes. 
In Sec.~\ref{sec:TDA}, we consider the template space 
$\{f_c,\,Q\}$ for the white noise case analytically and develop
an efficient tiling method. 
In Sec.~\ref{sec:TN}, 
by using a fitting curve for the TAMA DT 7 noise spectrum,
we show that our tiling method developed in the case of white noise
is valid even in the case of colored noise. 
Sec.~\ref{sec:Dis} is devoted to summary and discussion.
In Appendix A, we discuss simpler cases when only cosine or sine 
part of the ringdown waves are considered. 
In Appendix B, we discuss the number of templates in the case 
when the mass of the black hole is known.

%%%%%%%%%%%%%%%%%%%%%%%%%%%%%%%%%%%
%%%%%%%%%%%%%%%%%%%%%%%%%%%%%%%%%%%
\section{Quasi-normal ringing modes}\label{sec:FQM}
%%%%%%%%%%%%%%%%%%%%%%%%%%%%%%%%%%%
%%%%%%%%%%%%%%%%%%%%%%%%%%%%%%%%%%%

A black hole is characterized by its mass $M$ and 
spin angular momentum $J$.
Here we use the dimensionless spin parameter $a=J/M^2$ that takes
a value in the range $[0,1)$ with $a=0$ corresponding to
a Schwarzschild black hole and $a=1$ to extreme Kerr black hole.
Quasi-normal modes of a black hole are complex frequency solutions
of the Teukolsky wave equation that satisfy purely outgoing-wave
boundary condition at infinity and ingoing-wave at horizon
with vanishing incoming-wave amplitude. 
For fixed spheroidal harmonic indices $(\ell,m)$,
there are infinite number of quasi-normal modes.
They are assigned with an index $n$ with the order
of the magnitude of the imaginary part, i.e., 
the $n=1$ mode has the smallest imaginary part 
(or the largest quality factor).
We call it the least-damped (fundamental) mode.
It is known that the imaginary part of
the $\ell=m=2$ least-damped mode
is the smallest of all the quasi-normal modes, and 
results of black hole perturbation calculations as well
as numerical relativity simulations strongly suggest that
the ringdown waves are dominated by this
$\ell=m=2$ least-damped mode unless the spin parameter $a$
is extremely close to unity. 
Hence, we focus on this mode.

For the $\ell=m=2$ least-damped mode, analytical fitting formulas
for the central frequency and quality factor 
for a black hole of mass $M$ and dimensionless
spin $a$ were found by Echeverria as
\begin{eqnarray}
f_c &\simeq& 32{\rm kHz}\,[1-0.63(1-a)^{0.3}]
\left({M \over M_{\odot}}\right)^{-1} \,,
\label{eq:fcMa} 
\\
Q &\simeq& 2.0(1-a)^{-0.45} \,.
\label{eq:QMa}
\end{eqnarray}
The ringdown waveform is expressed as 
\begin{eqnarray}
h(f_c,\,Q,\,t_0,\,\phi_0;\,t) = \cases{
e^{ - \pi \,f_c\,(t-t_0)/Q}\,\cos(2\,\pi \,f_c\,(t-t_0)-\phi_0) 
& for 
$t \geq t_0$ \,,
\cr 
0 & for $t < t_0$ \,,
\cr}
\label{eq:RDwave}
\end{eqnarray}
where $t_0$ and $\phi_0$ are the initial time and phase of
the ringdown wave, respectively. 

%%%%%%%%%%%%%%%%%%%%%%%%%%%%%%%%%%%
%%%%%%%%%%%%%%%%%%%%%%%%%%%%%%%%%%%
\section{Template space}\label{sec:TDA}
%%%%%%%%%%%%%%%%%%%%%%%%%%%%%%%%%%%
%%%%%%%%%%%%%%%%%%%%%%%%%%%%%%%%%%%

In this section, we develop an efficient technique of template spacing 
which can be used for matched filtering of the 
quasi-normal ringing waveforms. 
Here, the detector noise is assumed to be white noise to make it possible
to deal with the problem analytically. 
The case of the colored noise is discussed in the next section. 

%%%%%%%%%%%%%%%%%%%%%%%%%%%%%%%%%%%
\subsection{Distance function}
%%%%%%%%%%%%%%%%%%%%%%%%%%%%%%%%%%%

We have temporarily set the amplitude to unity for simplicity 
in Eq.~(\ref{eq:RDwave}). 
Note that the knowledge of the amplitude is not necessary
for the template spacing in matched filtering.
For $t \geq t_0$, 
the ringdown wave (\ref{eq:RDwave}) is divided into two parts. 
\begin{eqnarray}
h(f_c,\,Q,\,t_0,\,\phi_0;\,t) &=& h_c(f_c,\,Q,\,t_0;\,t)\cos \phi_0 
+ h_s(f_c,\,Q,\,t_0;\,t)\sin \phi_0 \,, 
\end{eqnarray}
where
\begin{eqnarray}
h_c(f_c,\,Q,\,t_0;\,t) &=& 
e^{ - \frac {\pi \,f_c\,(t-t_0)}{Q}}\,\cos(2\,\pi \,f_c\,(t-t_0)) \,,
\\
h_s(f_c,\,Q,\,t_0;\,t) &=& 
e^{ - \frac {\pi \,f_c\,(t-t_0)}{Q}}\,\sin(2\,\pi \,f_c\,(t-t_0)) \,.
\end{eqnarray}

Performing the Fourier transformation 
$\tilde{h}_{c/s}(f)=\int_{-\infty}^{\infty} dt\,e^{2\pi i f t}h_{c/s}(t)$, 
we obtain the waveform in the frequency domain as 
\begin{eqnarray}
\tilde{h}_c(f_c,\,Q,\,t_0;\,f) &=& {\displaystyle \frac 
{( f_c - 2\,i\,f\,Q)\,Q \,e^{2\,i\,\pi\,f\,t_0}}
{\pi(  2\,f_c\,Q - i\,f_c - 2\,f\,Q)\,(2\,f_c\,Q + i\,f_c 
+ 2\,f\,Q)}} \,, \label{eq:FHC} \\ 
\tilde{h}_s(f_c,\,Q,\,t_0;\,f) &=& {\displaystyle \frac 
{2\,f_c\,Q^2\,\,e^{2\,i\,\pi\,f\,t_0}}
{\pi(  2\,f_c\,Q - i\,f_c - 2\,f\,Q)\,(2\,f_c\,Q + i\,f_c 
+ 2\,f\,Q)}} \,. \label{eq:FHS} 
\end{eqnarray}
The waveform in the time domain is real, 
so the following relation is satisfied. 
\begin{eqnarray}
\tilde{h}^*(f)=\tilde{h}(-f) \,,
\end{eqnarray}
where the star (${~}^*$) denotes the complex conjugation. 

Here, we introduce the inner product between two functions as 
\begin{eqnarray}
(a,\,b) = \int_{-f_{\rm max}}^{f_{\rm max}} df \,\tilde{a}(f)
\tilde{b}^*(f) \,,
\end{eqnarray}
where $f_{\rm max}$ is the maximum frequency we take into account in 
the analysis. In the actual data analysis, it is equal to or less than 
the half of the sampling frequency of data. 

In the matched filtering, 
we calculate the inner product between the template $h$ and the signal $x$ defined by $(x,h)$. 
We first introduce the normalized template.
We define the normalization constants as
\begin{eqnarray}
N_c(f_c,\,Q,\,t_0) &=& 
(\tilde{h}_c(f_c,\,Q,\,t_0),\tilde{h}_c(f_c,\,Q,\,t_0))
\nonumber \\ 
&=& {\displaystyle \frac {1}{2}} \,{\displaystyle \frac 
{(2\,Q^{2} + 1)\,Q}
{\pi \,(4\,Q^{2} + 1)\,f_c}} \,, \\
N_s(f_c,\,Q,\,t_0) &=& 
(\tilde{h}_s(f_c,\,Q,\,t_0),\tilde{h}_s(f_c,\,Q,\,t_0))
\nonumber \\ 
&=& {\displaystyle \frac 
{Q^3}
{\pi \,(4\,Q^{2} + 1)\,f_c}} \,,
\end{eqnarray}
when we set $f_{\rm max}=\infty$.
The normalized templates $\hat{h}_{c/s}(f_c,\,Q,\,t_0;\,f)$ are given by
\begin{eqnarray}
\hat{h}_{c/s}(f_c,\,Q,\,t_0;\,f) 
= {1 \over \sqrt{N_{c/s}(f_c,\,Q,\,t_0)}} \tilde{h}_{c/s}(f_c,\,Q,\,t_0;\,f) \,.
\end{eqnarray}
We then consider $\hat{h}=\hat{h}_c\cos\phi_0+\hat{h}_s\sin\phi_0$ as a template. 
We note that the $h_c$ and $h_s$ are not orthogonal. Their inner product is given by 
\begin{eqnarray}
(\hat{h}_c(f_c,\,Q,\,t_0),\hat{h}_s(f_c,\,Q,\,t_0)) 
&=& {1 \over \sqrt{2\,(2\,Q^2+1)}} 
\nonumber \\ &=:& c(f_c,\,Q,\,t_0) \,,
\end{eqnarray}
when $f_{\rm max}=\infty$.
In this case, the maximization of $(x,\hat{h})$ over the phase $\phi_0$ can be carried out 
analytically to yield \cite{Moh}
\begin{eqnarray}
\Lambda(x) &\equiv& \max_{\phi_0} (x,\hat{h}) \nonumber\\
&=& {(x,\,\hat{h}_c(f_c,\,Q,\,t_0))^2
+(x,\,\hat{h}_s(f_c,\,Q,\,t_0))^2
-2\,c(f_c,\,Q,\,t_0)(x,\,\hat{h}_c(f_c,\,Q,\,t_0))(x,\,\hat{h}_s(f_c,\,Q,\,t_0))
\over 1-c(f_c,\,Q,\,t_0)^2} \,,
\label{eq:LAM}
\end{eqnarray}
In the following, we consider the 3-dimensional template space 
$\{f_c,\,Q,\,t_0\}$. 

Here we consider the match $C(d f_c,\,dQ,\,dt_0)$
between the template with $(f_c,\,Q,t_0)$ and the normalized signal 
having slightly different sets of the parameters 
$(f_c+d f_c,\,Q+dQ,\,t_0+dt_0)$. 
Since we have already maximized over the phase $\phi_0$ in $\Lambda$,
without loosing generality, 
we only need to consider $\hat{h}_c$ for the signal. 
Then, the match is defined by 
\begin{eqnarray}
C(d f_c,\,dQ,\,dt_0) &=& \Lambda(\hat{h}_c(f_c+df_c,\,Q+dQ,\,t_0+dt_0))
\nonumber \\ &=& 
1- 
{\displaystyle \frac {2\,Q^4}{(2\,Q^{2} + 1)\,f_c^{2}}} \,df_c^{2} 
- 
{\displaystyle \frac {2\,Q^{2}\,(4\,Q^2+5)}
{(4\,Q^{2} + 1)^{2}(2\,Q^{2} + 1)}} \,dQ^{2}
\nonumber \\ &&
\mbox{} -
{\displaystyle \frac {2\,Q^{2}}
{f_c\,(4\,Q^{2} + 1)(2\,Q^{2} + 1)}} \,df_c\,dQ
\nonumber \\ &&
\mbox{} + 
{\displaystyle \frac {4\,\pi\,f_c\,(1+4\,Q^2)f_{\rm max}}
{Q\,(2\,Q^{2} + 1)}} \,dt_0^{2} 
+{\displaystyle \frac {\pi\,f_c}
{2\,Q^{2} + 1}} \,dt_0\,dQ
\,.
\end{eqnarray}
where we have used the approximation $f_{\rm max}\gg f_c$, and 
only the leading term in $f_{\rm max}/f_c$ are shown.  
We will find that the dependence of $f_{\rm max}$ will be 
cancelled out and does not appear in the final result. 
The inequality $C(d f_c,\,dQ,\,dt_0) \leq 1$ means that 
there will be a loss of signal-to-noise ratio unless the actual parameters of
a gravitational wave signal fall exactly onto one of the templates.

The smaller the match $C$ is, the larger the
distance is between the two signals in the template space.
Therefore, we define the metric in the template space by $ds^2_{(3)}=1-C$ \cite{owen},
that is,
\begin{eqnarray}
ds^2_{(3)} &=& g^{(3)}_{ij}dx^{i}dx^{j} 
\nonumber \\ &=& 
{\displaystyle \frac {2\,Q^4}{(2\,Q^{2} + 1)\,f_c^{2}}} \,df_c^{2} 
+ {\displaystyle \frac {2\,Q^{2}\,(4\,Q^2+5)}
{(4\,Q^{2} + 1)^{2}(2\,Q^{2} + 1)}} \,dQ^{2}
\nonumber \\ &&
\mbox{} -
{\displaystyle \frac {2\,Q^{3}}
{f_c\,(4\,Q^{2} + 1)(2\,Q^{2} + 1)}} \,df_c\,dQ
\nonumber \\ &&
\mbox{} + 
{\displaystyle \frac {4\,\pi\,f_c\,(1+4\,Q^2)f_{\rm max}}
{Q\,(2\,Q^{2} + 1)}} \,dt_0^{2} 
+{\displaystyle \frac {\pi\,f_c}
{2\,Q^{2} + 1}} \,dt_0\,dQ \,.
\end{eqnarray}

%%%%%%%%%%%%%%%%%%%%%%%%%%%%%%%%%%%
\subsection{Projection to 2-dimensional template space}\label{subsec:P2}
%%%%%%%%%%%%%%%%%%%%%%%%%%%%%%%%%%%

Now to maximize the match 
with respect to $dt_0$, we consider a projection of the distance function
into two dimensions spanned by $f_c$ and $Q$. 
Namely, we project $g^{(3)}_{ij}$ to a two dimensional subspace 
orthogonal to the $t_0$-axis \cite{owen} as
%We consider the projection of the distance function in
%the three dimensional space into two dimension. 
%In order to take the minimum of the distance function $ds^2_{(3)}$ 
%with respect to $dt_0$, 
%we project $g^{(3)}_{ij}$ on the 
%two dimensional template space orthogonal to $t_0$ \cite{owen} as 
\begin{eqnarray}
g^{(2)}_{IJ} = g^{(3)}_{IJ} 
- {g^{(3)}_{It_0}g^{(3)}_{Jt_0}\over g^{(3)}_{t_0 t_0}} \,, 
\label{eq:3to2}
\end{eqnarray}
where the indices $I,\,J$ are $\{f_c,\,Q\}$.
So, we find 
\begin{eqnarray}
ds^2_{(2)} &=& g^{(2)}_{IJ}dx^{I}dx^{J} 
\nonumber \\ &=& 
{\displaystyle \frac {2\,Q^4}{(2\,Q^{2} + 1)\,f_c^{2}}} \,df_c^{2} 
+ {\displaystyle \frac {2\,Q^{2}\,(4\,Q^2+5)}
{(4\,Q^{2} + 1)^{2}(2\,Q^{2} + 1)}} \,dQ^{2}
\nonumber \\ &&
\mbox{} -
{\displaystyle \frac {2\,Q^{3}}
{f_c\,(4\,Q^{2} + 1)(2\,Q^{2} + 1)}} \,df_c\,dQ 
\,,
\label{eq:ds2}
\end{eqnarray}
where we have taken the limit $f_{\rm max} \rightarrow \infty$. 

It is noted that the cross term $df_c\,dQ$ arises here. 
Later, we discuss a coordinate transformation in the template
space that removes the cross term, in order to
make our analysis of the template spacing 
and the error estimation easier.
We also note that the dependence of the metric on $f_c$ can be
eliminated by the simple coordinate transformation
$f_c\to\ln f_c$. 

It is also noted that the smaller the volume element of the metric is
the fewer the number of required filters is to cover the
template space.

The metric in the case of only $h_c$ or $h_s$ as templates 
is discussed in Appendix~\ref{app:NPC}. 

%%%%%%%%%%%%%%%%%%%%%%%%%%%%%%%%%%%
\subsection{Diagonalization of the metric}
%%%%%%%%%%%%%%%%%%%%%%%%%%%%%%%%%%%

Let us perform the coordinate transformation by which 
the two dimensional metric~(\ref{eq:ds2}) is transformed to 
a diagonal, conformally flat metric. 

We start from the coordinate transformation that
removes the frequency dependence in the metric. 
We set 
\begin{eqnarray}
dF = \frac{d f_c}{f_c} \,, \label{eq:defdF}
\end{eqnarray}
which gives
\begin{eqnarray}
ds^2 &=&  g_{FF} \,dF^{2} 
+ g_{QQ} \,dQ^{2}
+2\,g_{FQ} \,dF\,dQ 
\nonumber\\
 &=& {\displaystyle \frac {2\,Q^4}{(2\,Q^{2} + 1)}} \,dF^{2} 
+ {\displaystyle \frac {2\,Q^{2}\,(4\,Q^2+5)}
{(4\,Q^{2} + 1)^{2}(2\,Q^{2} + 1)}} \,dQ^{2}
\nonumber \\ &&
\mbox{} -
{\displaystyle \frac {2\,Q^{3}}
{(4\,Q^{2} + 1)(2\,Q^{2} + 1)}} \,dF\,dQ  \,.
\label{eq:ds2NN}
\end{eqnarray}
The transformation that removes the off-diagonal element
is found by setting $F=X-u(Q)$ and
requiring
\begin{eqnarray}
g_{FF}\,u'(Q)-g_{FQ}=0\,.
\label{eq:udeF}
\end{eqnarray}
We find
\begin{eqnarray}
u= {1 \over 2} \ln \left(4+{1 \over Q^2}\right)\,,
\label{eq:usoL}
\end{eqnarray}
which gives
\begin{eqnarray}
ds^2
&=&g_{FF}\,dX^2
+\left(g_{FF}\,u'{}^2-2g_{FQ}\,u'+g_{QQ}\right)dQ^2 \nonumber\\
&=&g_{FF}\left(dX^2+\frac{g_{FF}\,g_{QQ}-g_{FQ}^2}{g_{FF}^2}dQ^2\right)\,.
\end{eqnarray}
Then we can perform a further coordinate transformation to make
the metric conformally flat. Namely, by the transformation
$Q\to Y$ defined by
\begin{eqnarray}
Y=\int_{Q}^\infty dQ'{\sqrt{\det g(Q')\over g_{FF}(Q')}}
=\int_{Q^2}^\infty dx\,
{\sqrt{x+1} \over x\,(4\,x+1)}
\quad(Q>0),
\label{eq:YdeF}
\end{eqnarray}
we obtain
\begin{eqnarray}
ds^2=\Omega(Y)\left(dX^2+dY^2\right)\,,
\label{eq:ds2coN}
\end{eqnarray}
where the conformal factor is given by
\begin{eqnarray}
\Omega(Y)=g_{FF}\bigl(Q(Y)\bigr).
\end{eqnarray}
Here $Q$ is now a function of $Y$ determined by inverting
Eq.~(\ref{eq:YdeF}).

Although the above coordinate transformation involves complicated
functions that may not be expressed in terms of elementary functions,
we find it is sufficient to use their large $Q$ expansion forms 
for $Q\geq2$.
Up to $O(1/Q^8)$ inclusive, we have
\begin{eqnarray}
X&=&F+u=F + \ln 2 + {1 \over 8}{1 \over Q^2}
-{1 \over 64}{1 \over Q^4}+{1 \over 384}{1 \over Q^6}
-{1 \over 2048}{1 \over Q^8} \,, \\ 
Y &=& 
{1 \over 2}{1 \over Q}
+{1 \over 24}{1 \over Q^3}-{3 \over 160}{1 \over Q^5}
+{1 \over 128}{1 \over Q^7}-{17 \over 4608}{1 \over Q^9} \,.
\end{eqnarray}
To the same accuracy, the inverse transformation becomes
\begin{eqnarray}
F &=& X - \ln 2 
-{1 \over 2}{Y^2}
+{7 \over 12}{Y^4}-{67 \over 45}{Y^6}
+{1769 \over 360}{Y^8} \,, 
\\ 
Q &=& {1 \over 2}{1 \over Y}
+{1 \over 6}Y-{37 \over 90}Y^3
+{166 \over 135}Y^5-{5917 \over 1350}Y^7 \,.
\end{eqnarray}
The conformal factor $\Omega(Y)$ is given by
\begin{eqnarray}
\Omega(Y) &=& {1 \over 4}{1 \over Y^2}-{1 \over 3}
+{37 \over 60}{Y^2}-{85 \over 54}{Y^4}
+{13069 \over 2700}{Y^6} \,.
\end{eqnarray}
When $Q=2$, the errors induced by the above expansion are found to be
$\sim 0.1\,\%$. This is accurate enough for our purpose
as long as we allow the SNR loss, $ds^2_{\rm max}$, of a few percent.

%%%%%%%%%%%%%%%%%%%%%%%%%%%%%%%%%%%
\subsection{Tiling method: basis}
%%%%%%%%%%%%%%%%%%%%%%%%%%%%%%%%%%%

In the previous subsection, 
we have derived the simple, conformally flat metric~(\ref{eq:ds2coN})
for the template space. Here, using this metric, we formulate 
a tiling algorithm which is not only efficient but also quite simple.

To develop such a method, we note the following.
Because of the conformal flatness, the contour of the fixed maximum
distance $ds^2=ds^2_{\rm max}$ centered at a point on the
$(X,Y)$-plane is a circle for sufficiently small $ds^2_{\rm max}$.
Furthermore, along a line of $Y=$constant, $\Omega(Y)$ is constant.
Thus, choosing first an appropriate $Y=$constant line,
say $Y=q_1$, we may place circles of the same radius with their
centers located along the line $Y=q_1$ to cover a region
surrounding that line. Then, if we find an algorithm to 
place circles along the $Y=q_1$ line and another algorithm to
choose the next $Y=$constant line, say $Y=q_2$, to be covered in an
appropriate way, we can repeat this tiling procedure to
cover the whole template space.

Let us assume that the template space to be tiled is
a rectangle given by $F_{\rm min}\leq F\leq F_{\rm max}$
and $Q_{\rm min}\leq Q\leq Q_{\rm max}$. In the $(X,Y)$ 
coordinates, this rectangle is mapped to the region
bounded by the two $Y=$constant lines
corresponding to $Q=Q_{\rm min}$ and $Q=Q_{\rm max}$, 
which we denote by $Y=Y_0$ and $Y_{\rm M}$, respectively,
and the two lines $X=v_{\rm min}(Y)$ and $X=v_{\rm max}(Y)$
corresponding to $F=F_{\rm min}$ and $F=F_{\rm max}$, respectively.
Note that $Y_0>Y_{\rm M}$ since large $Y$ corresponds to small $Q$.

First, we construct a method to determine 
the spacing of the circles along each $Y=$constant line. 
Let us consider the line $Y=q$ and
place two circles with radius $r$
centered at $(p,\,q)$ and $(p+\Delta p,\,q)$,
\begin{eqnarray}
(X-p)^2 + (Y-q)^2 = r^2 \,,
\quad
(X-p-\Delta p)^2 + (Y-q)^2 = r^2 \,.
\end{eqnarray}
We assume $\Delta p<2r$ so that the two circles intersect at 
the two points, $\left(p+{\Delta p/2}\,,\,q\pm d(r;\,p)\right)$,
where $d(r;\,p)$ is the distance to each intersecting point
from the line $Y=q$ (see Fig.\ref{fig:umo12}), given by
\begin{eqnarray}
d(r;\,\Delta p) = \sqrt{r^2-{\Delta p^2 \over 4}} \,.
\end{eqnarray}
Our purpose is to tile the template space by the smallest 
possible number of filters. In order to do so, we choose the
parameter $p$ in such a way that the area defined by 
$S=\Delta p\,d(r;\,\Delta p)$ is maximized, i.e., 
\begin{eqnarray}
\Delta p &=& \sqrt{2}\,r \,, \\
d &=& {r \over \sqrt{2}} \,.
\end{eqnarray}
The radius $r$ is determined by 
the value of $ds_{\rm max}^2$ and $q$ as
\begin{eqnarray}
r^2={ds_{\rm max}^2\over\Omega(q)}\,.
\label{eq:rdet}
\end{eqnarray}
In this way, we tile the region that covers the line $Y=q$.

\begin{figure}[ht]
\begin{center}
\epsfxsize=8cm
\hspace{-1cm}\epsfbox{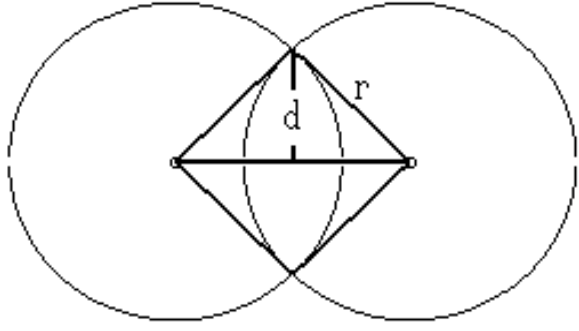}
\caption{Definition of $d$ and $r$.}
\label{fig:umo12}
\end{center}
\end{figure}

To choose the first line to be covered, we start from
the point on the $(X,Y)$-plane
corresponding to $(F,Q)=(F_{\rm max},\,Q_{\rm min})$,
that is, $(X,Y)=(X_0,Y_0)$ where $X_0=v_{\rm max}(Y_0)$. 
Then we choose the first line $Y=q_1$ and the radius $r_1$
so that the point $(X_0,Y_0)$ is just on the edge of the first
circle and an intersecting point of the first and second circles
lies on the line $Y=Y_0$ as Fig.~\ref{fig:umo1}. 
This is achieved if the center of the first circle is at 
\begin{eqnarray}
(p_1,\,q_1)=
\left({X_0-r_1/\sqrt{2}\,,\,Y_0-r_1/\sqrt{2}}\right)\,,
\end{eqnarray}
with the radius determined by solving Eq.~(\ref{eq:rdet}),
which reads in the present case,
\begin{eqnarray}
ds_{\rm max}^2=r_1^2\,\Omega(Y_0-r_1/\sqrt{2})\,.
\label{eq:r1eq}
\end{eqnarray}

\begin{figure}[ht]
\begin{center}
\epsfxsize=8cm
\hspace{-1cm}\epsfbox{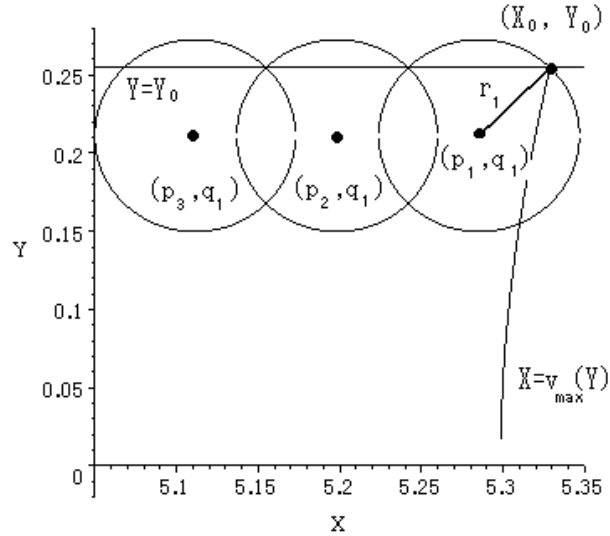}
\caption{Choosing the first line on the $(X,Y)$-plane.
}
\label{fig:umo1}
\end{center}
\end{figure}
Then the center of the $n$-th circle is at
\begin{eqnarray}
(p_n,\,q_1)=
\left({X_0-(2n-1)\,r_1/\sqrt{2}\,,\,Y_0-r_1/\sqrt{2}}\right)\,,
\end{eqnarray}
and the number of circles needed to cover the line $Y=q_1$
is given by
\begin{eqnarray}
N_1=\left[{L_1\over \sqrt{2}\,r_1}\right],
\label{eq:N1}
\end{eqnarray}
where $[x]$ denotes the maximum integer smaller than $x+1$
and $L_1$ is the coordinate length of $X$ to be covered,
i.e., 
\begin{eqnarray}
L_1 &=& v_{\rm max}(Y_0)-v_{\rm min}(Y_0) 
\nonumber \\ 
&=& F_{\rm max} - F_{\rm min} 
\nonumber \\ &=:& L \,.
\end{eqnarray}

Once the covering of the first line is done, 
the second $Y=$constant line is chosen as follows.
Let $Y_1=Y_0-\sqrt{2}r_1$ and let $(X_1,Y_1)$ be the intersecting 
point of the lines $Y=Y_1$ and $X=v_{\rm max}(Y)$,
i.e., $(X_1,Y_1)=(v_{\rm max}(Y_1),Y_1)$.
We choose this point as the starting point for the covering
of the second line as Fig.~\ref{fig:umo3}. 
\begin{figure}[ht]
\begin{center}
\epsfxsize=8cm
\hspace{-1cm}\epsfbox{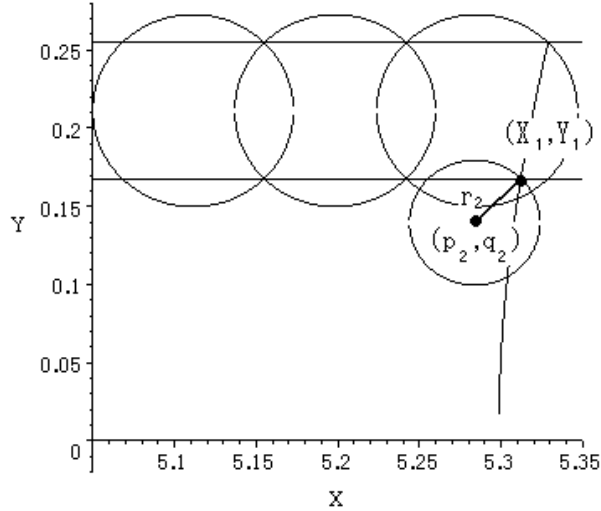}
\caption{Covering of the second line.
}
\label{fig:umo3}
\end{center}
\end{figure}
That is, the center of the
first circle $(p_2,q_2)$ on the second line $Y=q_2$ is 
\begin{eqnarray}
(p_2,\,q_2)=
\left({X_1-r_2/\sqrt{2}\,,\,Y_1-r_2/\sqrt{2}}\right)\,,
\end{eqnarray}
with the radius $r_2$ determined again by Eq.~(\ref{eq:rdet}).
Then the second line is covered by the same procedure
we took for the first line. 
We repeat this procedure until we tile the whole template
space we need to cover.

With this tiling procedure, the total number of templates
is given as follows.
Generalizing Eq.~(\ref{eq:N1}),
the number of templates for the $i$-th $Y=$constant line 
($Y=q_i$) is
\begin{eqnarray}
N_i=\left[{L \over \sqrt{2}\,r_i}\right]\,,
\end{eqnarray}
where 
$r_i$ is determined by
\begin{eqnarray}
ds_{\rm max}^2=r_i^2\,\Omega(Y_{i-1}-r_i/\sqrt{2})\,.
\end{eqnarray}
The number of $Y=$constant lines necessary to cover
the template space is determined by the minimum integer
$\nu$ that satisfies the inequality,
\begin{eqnarray}
\sum_{i=1}^{\nu}\sqrt{2}\,r_i \geq Y_0-Y_{\rm M}\,.
\end{eqnarray}
And the total number of templates is 
\begin{eqnarray}
{\cal N} &=& \sum_{i=1}^{\nu} N_i \,.
\end{eqnarray}

%%%%%%%%%%%%%%%%%%%%%%%%%%%%%%%%%%%
\subsection{Tiling method: application}
%%%%%%%%%%%%%%%%%%%%%%%%%%%%%%%%%%%

Let us apply the method developed in the previous
subsection to the case of the parameter space $(f_c,\,Q)$ 
which has the range, 
\begin{eqnarray}
10^2{\rm Hz} \leq &f_c & \leq 10^4{\rm Hz} \,, 
\nonumber \\ 
2 \leq &Q& \leq 20 \,.
\end{eqnarray}
We set $ds^2_{\rm max}=0.02$. This choice is made
in order to make the SNR loss to be $\sim 3\,\%$ 
in the presence of colored noise as discussed
in the next section.
Using Eq.(\ref{eq:defdF}), we set
$F=\ln(f_c/100{\rm Hz})$. 
The above range corresponds to 
\begin{eqnarray}
6.93\times 10^{-1} \leq &X& \leq 5.33 \,, \nonumber \\ 
2.50\times 10^{-2} \leq &Y& \leq 2.55\times 10^{-1} \,.
\end{eqnarray}
in the $(X,\,Y)$-space. 
\begin{table}[ht]
 \begin{center}
\renewcommand{\arraystretch}{1.8}
  \begin{tabular}{|c|c|}
\hline
$(f_c,\,Q)$ & $(X,\,Y)$  \\ 
\hline
$10^2$Hz,\,2.0 & 7.234594008 $10^{-1}$,\,2.546762255 $10^{-1}$ \\ 
\hline
$10^2$Hz,\,20.0 & 6.934595830 $10^{-1}$,\,2.500520248 $10^{-2}$ \\
\hline
$10^3$Hz,\,2.0 & 3.026044494,\,2.546762255 $10^{-1}$ \\ 
\hline
$10^3$Hz,\,20.0 & 2.996044676,\,2.500520248 $10^{-2}$ \\ 
\hline
$10^4$Hz,\,2.0 & 5.328629588,\,2.546762255 $10^{-1}$ \\ 
\hline
$10^4$Hz,\,20.0 & 5.298629769,\,2.500520248 $10^{-2}$ \\ 
\hline
  \end{tabular}
 \end{center} 
\caption{The relation between the two coordinates}
\label{table:trans}
\end{table}

Following the procedure described in the previous subsection,
we choose the starting point $(X_0,\,Y_0)$ which 
corresponds to $(f_c,\,Q)=(10^4{\rm Hz},2.0)$. 
The radius $r_1$ is determined by Eq.~(\ref{eq:r1eq}), or
\begin{eqnarray}
0.02=r_1^2\,\Omega(Y_0-r_1/\sqrt{2})\,.
\end{eqnarray}
Then all the parameters for our tiling method are determined.
In Table~\ref{table:param}, we summarize the 
tiling parameters.
\begin{table}[ht]
 \begin{center}
\renewcommand{\arraystretch}{1.8}
  \begin{tabular}{|c||c|c|c|c|}
\hline
$i$ & $(p_i,\,q_i)$ & $(X_i,\,Y_i)$ & $r_i$ & $N_i$ \\ 
\hline
0 & & 5.328629588,\, 2.546762255 $10^{-1}$ & &  \\ 
\hline
1 & 5.285172504,\, 2.112191414 $10^{-1}$ & 5.311957476,\, 1.677620573 $10^{-1}$ 
& 6.145759769 $10^{-2}$ & 53 \\
\hline
2 & 5.283700230,\, 1.395048115 $10^{-1}$ & 5.304418737,\, 1.112475658 $10^{-1}$ 
& 3.996178019 $10^{-2}$ & 82 \\ 
\hline
3 & 5.285789828,\, 9.261865694 $10^{-2}$ & 5.301037366,\, 7.398974808 $10^{-2}$ 
& 2.634525556 $10^{-2}$ & 124 \\ 
\hline
4 & 5.288679812,\, 6.163219420 $10^{-2}$ & 5.299527944,\, 4.927464032 $10^{-2}$ 
& 1.747622029 $10^{-2}$ & 187 \\ 
\hline
5 & 5.291307826,\, 4.105452276 $10^{-2}$ & 5.298855740,\, 3.283440519 $10^{-2}$
& 1.162500174 $10^{-2}$ & 281 \\ 
\hline
6 & 5.293381065,\, 2.735973019 $10^{-2}$ & 
& 7.742359638 $10^{-3}$ & 421 \\ 
\hline
  \end{tabular}
 \end{center}
\caption{The parameters for the tiling of the template space are summarized. 
$(X_i,\,Y_i)$ denotes the starting point of the $i$-th template spacing
along the line $y=q_i$. 
$(p_i,\,q_i)$ denotes the center of the first circle along the $i$-th line
$y=q_i$, $r_i$ is the radius of the circle, and $N_i$ is
the number of circles necessary to cover the $i$-th line.
}
\label{table:param}
\end{table}

It is noted that the number $\nu$ of the $Y=$constant lines is 
very small, 
\begin{eqnarray}
\nu=6 \,.
\end{eqnarray}
The total number of templates is calculated to be ${\cal N}=1148$ 
(${\cal N}=1780$ for $Q_{\rm max}=34.3$). 
In Fig.~\ref{fig:umeta1}, we show the tiling of the template space 
in the $(X,\,Y)$ coordinates.
\begin{figure}[ht]
\center
\epsfxsize=8cm
\hspace{-1cm}\epsfbox{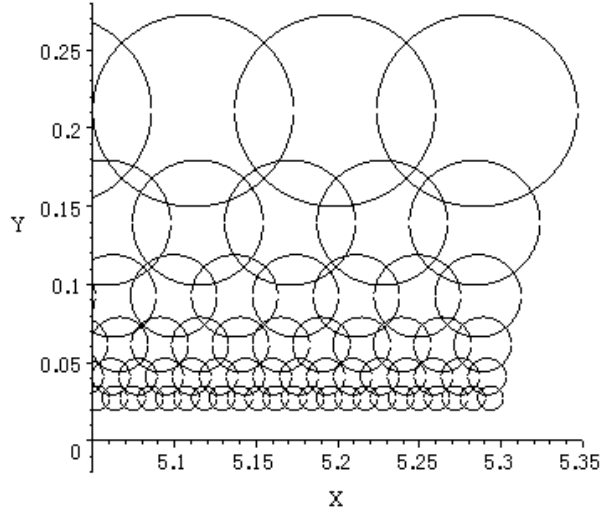}
\caption{A part of the tiling of the template space
in the $(X,Y)$ coordinates. 
Templates are taken at the centers of the circles. 
}
\label{fig:umeta1}
\end{figure}
\begin{figure}[ht]
\begin{center}
\epsfxsize=15cm
\hspace{-1cm}\epsfbox{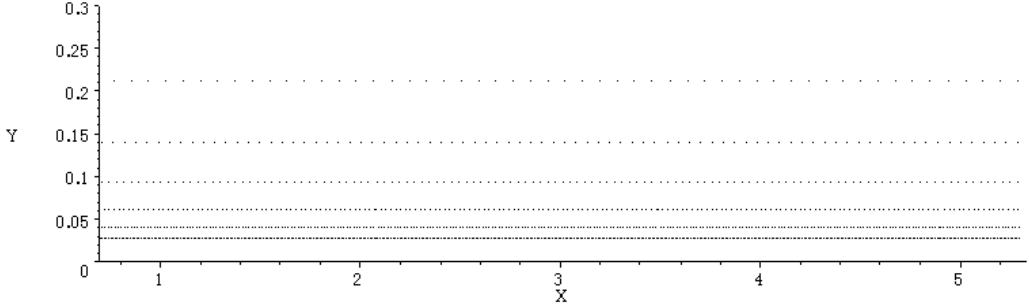}
\caption{The template points on the $(X,Y)$-plane.
It is noted that we may put the templates with the same 
interval along the $X$ direction. 
}
\label{fig:ume1}
\end{center}
\end{figure}
The tiling of the template space in the original coordinates $(f_c,\,Q)$
is shown in Fig.~\ref{fig:umeta2}.
\begin{figure}[ht]
\center
\epsfxsize=10cm
\hspace{-1cm}\epsfbox{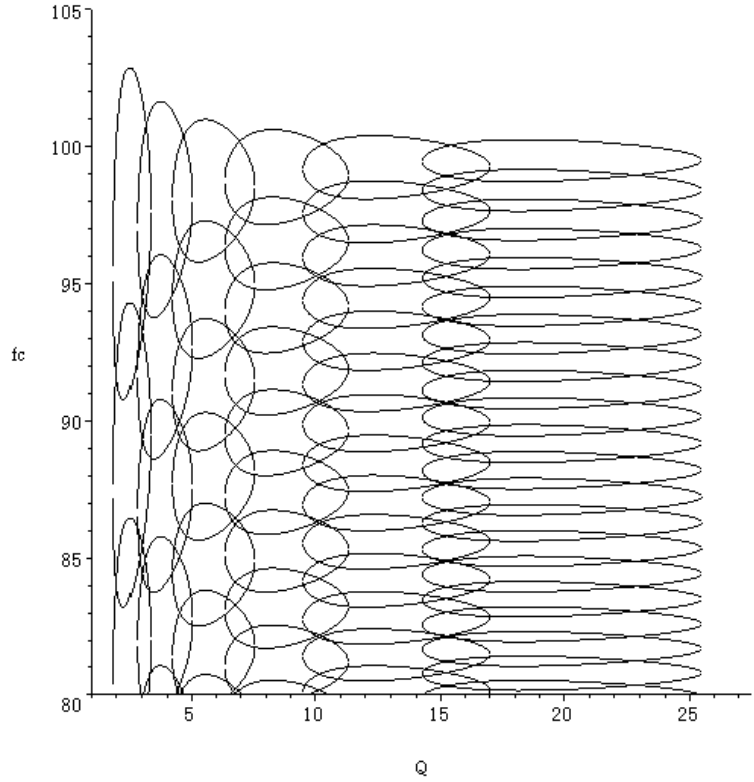}
\caption{A part of the tiling of the template space
in the original coordinates $(f_c,\,Q)$ with $f_c$ measured in
units of $100{\rm Hz}$.
Note that the contour of $ds^2_{\rm max}=0.02$ 
for each template is warped. 
}
\label{fig:umeta2}
\end{figure}
\begin{figure}[ht]
\begin{center}
\epsfxsize=15cm
\hspace{-1cm}\epsfbox{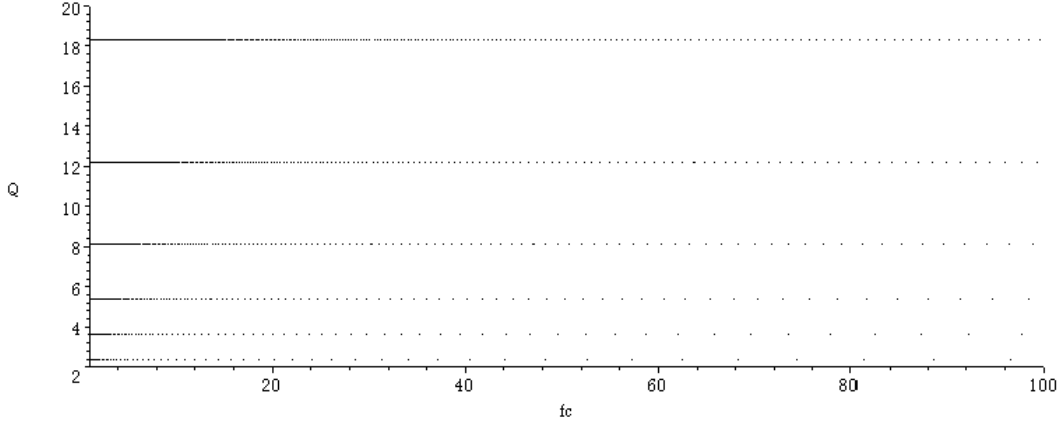}
\caption{The template points on the original coordinates 
$(f_c,\,Q)$. The smaller the value of $f_c$ is, the larger 
the number of templates is needed. 
}
\label{fig:ume2}
\end{center}
\end{figure}

In order to see how effective this tiling method is, 
we calculate the ratio between the sum of the areas
of all the circles ${\cal S}_{\rm cir}$ 
and the area to be covered ${\cal S}_{\rm par}$.
We find
\begin{eqnarray}
\eta &=& {{\cal S}_{\rm cir} \over {\cal S}_{\rm par}} 
\nonumber \\ &=& 1.57 \,.
\end{eqnarray}
Here we have adopted $Q_{\rm max}=22.9$ which is slightly
larger than the pre-assigned value of $Q_{\rm max}=20$.
The former value corresponds to the value 
$Y=q_6-r_6/\sqrt{2}$, i.e., the value of the $Y=$constant line
marginally covered by the circles centered along the
line $Y=q_6$.
This ratio corresponds to $\eta_{\rm tot}$ in the paper by 
Arnaud et al.~\cite{arna}, in which they obtained
$\eta_{\rm tot}=2.12$ for their tiling method. 
Thus our tiling method is far more efficient (and simpler)
than that of Arnaud et al..

%%%%%%%%%%%%%%%%%%%%%%%%%%%%%%%%%%%
%%%%%%%%%%%%%%%%%%%%%%%%%%%%%%%%%%%
\section{Validity of tiling method in the case of colored noise spectrum}\label{sec:TN}
%%%%%%%%%%%%%%%%%%%%%%%%%%%%%%%%%%%
%%%%%%%%%%%%%%%%%%%%%%%%%%%%%%%%%%%

The tiling method discussed in the previous section
is based on an assumption that the noise is white.
We expect this assumption is very good because the ringing wave 
is rather narrow banded except the case $Q\sim 2$. 
In order to confirm this, we examine the effectiveness of 
the tiling method in the case of colored noise. 

As a model of detector's noise, 
we use a fitting curve of the one sided noise power spectrum of TAMA300,
which is given by 
\begin{eqnarray}
S_n(|f|)&=&\left({85 \over f}\right)^{63} 
+ {1\over 2}\,\left({220 \over f}\right)^{10} 
+ {1\over 9}\,\left({710 \over f}\right)^{3} 
+ {3 \over 20}
+ {1\over 20}\,\left({f \over 2000}\right)^2 
+ {1\over 5}\,\left({f \over 5500}\right)^6. 
\end{eqnarray}
This formula of the noise spectrum is obtained by fitting Fig. 3.4.1 
in TAMA REPORT 2002 \cite{TAMAreport} which is based on the spectrum
during Data Taking 7 in 2002, 
and is valid between $60$ Hz and $40000$ Hz. 
Here we have ignored the overall amplitude of $S_n$, 
because it does not affect the results. 

We prepare the template bank using the analytical method 
in the white noise case. The minimal match is assumed to be 0.98. 
We also generate signals which are normalized to unity. 
We then calculate the maximum of the match between the signal and templates.
When the signal is completely the same, the match becomes unity. 
However, we have the match less than unity due to mismatch between signal
and templates. If the match is always greater than 0.98, 
we have a justification to use the tiling method even in the case 
of the colored noise spectrum. 

We used 2500 signals. 
The range of the central frequency $f_c$ and the $Q$ value of the signals
are $100$Hz $\leq f_c \leq 10^4$Hz and $2 \leq Q \leq 20$ respectively. 
The value of $f_c$ and $Q$ are randomly given in this region. 

In Fig.~\ref{fig:effnum}, we show the number of signals
in terms of the value of the match. 
\begin{figure}[ht]
\center
\epsfxsize=10cm
\hspace{-1cm}\epsfbox{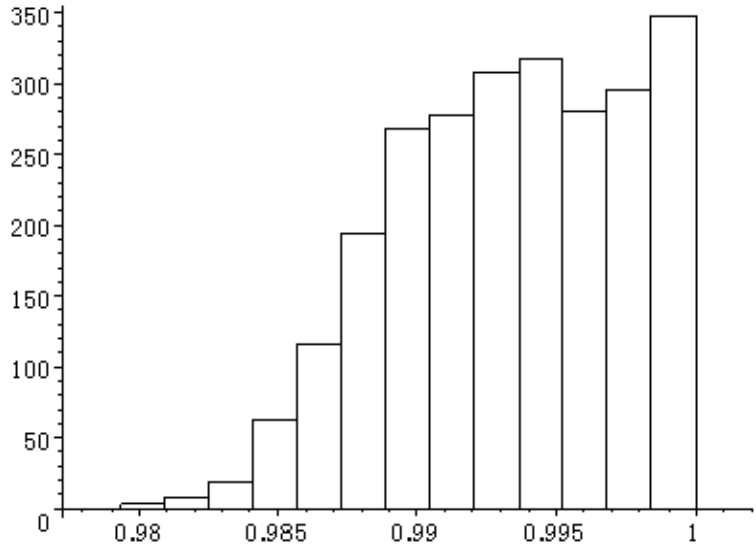}
\caption{
The number of signals 
in terms of the value of the match between the signal and templates. 
The bank of templates are determined 
assuming the minimum match 0.98 using the method in Section III. 
The mean value of the match is $0.993$.
}
\label{fig:effnum}
\end{figure}

From Fig.~\ref{fig:effnum}, 
we find that we can detect the most of the signals 
without losing $2\%$ of the signal-to-noise ratio 
by the tiling method assuming white noise. 
Thus, the tiling method, constructed analytically 
assuming white noise, 
is valid even in the case of the TAMA noise spectrum. 

%%%%%%%%%%%%%%%%%%%%%%%%%%%%%%%%%%%
%%%%%%%%%%%%%%%%%%%%%%%%%%%%%%%%%%%
\section{Discussion}\label{sec:Dis}
%%%%%%%%%%%%%%%%%%%%%%%%%%%%%%%%%%%
%%%%%%%%%%%%%%%%%%%%%%%%%%%%%%%%%%%

In this paper, we proposed a tiling method 
of the template space in the matched filtering search 
for the gravitational waves of black hole ringing. 

First, we discussed a tiling method assuming 
the detector noise is white. 
We found that the $1/Q$-expansion was useful in performing the coordinate 
transformation of the metric of the template space because of $Q \geq 2$. 
We also find an efficient tiling method of the template space 
which can been formulated analytically. 
When the range of $f_c$ and $Q$ is 
$10^2\leq f_c \leq 10^4$ Hz and $2\leq Q \leq 20$, 
and the minimal match is 0.97, 
the ratio $\eta$, between the sum of the areas of the equal match circles around 
the templates, ${\cal S}_{\rm cir}$, 
and the area to be covered, ${\cal S}_{\rm par}$,
is 1.57 which is much smaller than the value obtained before. 

Next, we have discuss the validity of 
this tiling method in the case of colored noise spectrum. 
As a model of realistic noise power spectrum, we used 
a fitting curve of the noise power spectrum of TAMA300 
during 2002. 
We prepare a template bank 
assuming the minimum match $98\%$. 
We then examined the loss of signal-to-noise ratio
of the signals detected in the template bank. We found that,
in most case, 
we do not lose the signal-to-noise ratio no more than $2\%$
which is expected from the pre-assigned minimum match. 
This shows that the tiling method can be used even 
in the case of colored noise spectrum. 

The tiling method should be tested using the real data of 
laser interferometers. 
A study in this direction 
using the data of TAMA300 is now progressing \cite{Tsune}. 

In the analysis using real interferometers' data, we have to 
treat the non-stationary, non-Gaussian noise. 
It is found \cite{TsuneKanda} that we observe even more fake events 
than in the case of inspiraling wave, 
since the ringing wave is usually much shorter than the inspiral waves 
and can easily be affected by short bursts. 
In this situation, we would need to introduce some methods
to remove such fake events without losing real ringing gravitational
wave signals. 
Coincidence analysis between several detectors 
would also be needed to reduce the fake event rate. 
Further, since the ringing waves may be excited 
by the detector itself, we may need to perform coincidence analysis
to reject such spurious events anyway. 
We will also work on this problem in the future.

%===========================%
\acknowledgments
%===========================%

We would like to thank  N. Kanda and Y. Tsunesada 
for useful discussions and comments. 
HN is supported by Research Fellowships of the
Japan Society for the Promotion of Science
for Young Scientists, No.~5919. 
This work is supported in part by Monbukagaku-sho Grant-in-Aid
for Scientific Research No.~14047214 and 12640269. 

%======================================%
%<<<<<<<<<<<<< APPENDIX >>>>>>>>>>>>>>>%
%======================================%
\begin{appendix}

%%%%%%%%%%%%%%%%%%%%%%%%%%%%%%%%%%%
\section{The metric in some special cases} \label{app:NPC}
%%%%%%%%%%%%%%%%%%%%%%%%%%%%%%%%%%%

In this appendix, we summarize 
the distance function in the case when we consider only
the cosine or sine part of the wave, ignoring the contribution 
of the phase and the initial time. 
We also consider the case when the detector noise is white.
Although the applicability of such special cases 
will be limited, we describe them here because there are still some 
cases when these formulas become useful. 

%%%%%%%%%%%%%%%%%%%%%%%%%%%%%%%%%%%
\subsection{Cosine wave case} \label{app:cos}
%%%%%%%%%%%%%%%%%%%%%%%%%%%%%%%%%%%

We consider the cosine waveform 
ignoring the initial time and phase of a ringdown wave;
\begin{eqnarray}
h(f_c,\,Q;\,t) = \cases{
e^{ - \pi \,f_c\,t/Q}\,\cos(2\,\pi \,f_c\,t) & for 
$t \geq 0$ \,,
\cr 
0 & for $t < 0$ \,.
\cr}
\label{eq:coswave}
\end{eqnarray}

Performing the Fourier transformation 
$\tilde{h}(f)=\int_{-\infty}^{\infty} dt\,e^{2\pi i f t}h(t)$, 
we obtain the waveform in the frequency domain as 
\begin{eqnarray}
\tilde{h}(f_c,\,Q;\,f) = {\displaystyle \frac {( f_c - 2\,i\,
f\,Q)\,
Q}{\pi(  2\,f_c\,Q - i\,f_c - 2\,f\,Q)\,(2\,f_c\,Q + i\,f_c 
+ 2\,f\,Q)}} \,.
\end{eqnarray}

In the following, we consider normalized waveforms.
The normalization constant is derived as
\begin{eqnarray}
N(f_c,\,Q) &=& \int_{-\infty}^{\infty} |\tilde{h}(f_c,\,Q;\,f)|^2 df 
\nonumber \\ 
&=& {\displaystyle \frac {1}{2}} \,{\displaystyle \frac 
{(2\,Q^{2} + 1)\,Q}
{\pi \,(4\,Q^{2} + 1)\,f_c}} \,.
\end{eqnarray}
With the above normalization constant, 
the normalized wave $\hat{h}(f_c,\,Q;\,f)$ is given by
\begin{eqnarray}
\hat{h}(f_c,\,Q;\,f) = {1 \over \sqrt{N(f_c,\,Q)}} \tilde{h}(f_c,\,Q;\,f) \,.
\end{eqnarray}

Now, we consider the correlation between the two normalized waves 
having slightly different sets of the parameters $(f_c,\,Q)$ and 
$(f_c+d f_c,\,Q+dQ)$,
\begin{eqnarray}
C(d f_c,\,dQ) &=& 
\int_{-\infty}^{\infty} df \,\hat{h}(f_c,\,Q;\,f)
\hat{h}^*(f_c+df_c,\,Q+dQ;\,f) 
\nonumber \\ &=& 
1- {\displaystyle \frac {1}{8}} \,
{\displaystyle \frac {16\,Q^4+6\,Q^{2} + 1}{(2\,Q^{2} + 1)\,f_c^{2}}} \,df_c^{2} 
- {\displaystyle \frac {1}{8}} \,
{\displaystyle \frac {64\,Q^8+128\,Q^{6} + 28\,Q^{4} + 1}
{Q^2\,(4\,Q^{2} + 1)^{2}(2\,Q^{2} + 1)^{2}}} \,dQ^{2}
\nonumber \\ &&
\mbox{}  + 
{\displaystyle \frac {1}{4}} \,
{\displaystyle \frac {(8\,Q^{4} + 2\,Q^{2}+ 1)}
{f_c\,Q\,(4\,Q^{2} + 1)(2\,Q^{2} + 1)}} \,df_c\,dQ
\,.
\end{eqnarray}
The inequality $C(d f_c,\,dQ) \leq 1$ means that 
there will be a loss of the signal to noise ratio unless the actual parameters of
a gravitational wave signal fall exactly onto one of the
templates.

The smaller the correlation $C$ is, the larger the
distance is between the two signals in the template space.
Therefore, we define the metric in the template space by $ds^2=1-C$,
that is,
\begin{eqnarray}
ds^2 &=& {\displaystyle \frac {1}{8}} \,
{\displaystyle \frac {16\,Q^4+6\,Q^{2} + 1}{(2\,Q^{2} + 1)\,f_c^{2}}} \,df_c^{2} 
+ {\displaystyle \frac {1}{8}} \,
{\displaystyle \frac {64\,Q^8+128\,Q^{6} + 28\,Q^{4} + 1}
{Q^2\,(4\,Q^{2} + 1)^{2}(2\,Q^{2} + 1)^{2}}} \,dQ^{2}
\nonumber \\ &&
\mbox{}  - 
{\displaystyle \frac {1}{4}} \,
{\displaystyle \frac {8\,Q^{4} +2\,Q^{2}+ 1}
{f_c\,Q\,(4\,Q^{2} + 1)(2\,Q^{2} + 1)}} \,df_c\,dQ
\label{eq:ds2c}
\,, \\
& \simeq & {Q^2 \over f_c^2} \,df_c^{2} 
+ {\displaystyle \frac {1}{8\,Q^2}} \,dQ^{2}
- {\displaystyle \frac {1}{4\,f_c\,Q}} \,df_c\,dQ \quad 
\mbox{for large}~Q.
\label{eq:ds2app}
\end{eqnarray}
Note that the dependence of the metric on $f_c$ can be
eliminated by the simple coordinate transformation
$f_c\to \ln f_c$.
The metric in the case of the sine-wave templates instead
of the cosine-wave templates (\ref{eq:coswave}) is 
calculate in the next subsection. 
For large $Q$, the two metrics coincide with each other,
implying that the phase effect is weak in the large $Q$ limit.

The number of filters, $N_{\rm f}$, 
which are required to achieve the detection efficiency determined
by the maximum allowable distance $ds^2_{\rm max}$
(i.e., the maximum allowable loss of SNR)
in the region
$f_{\rm min} \leq f \leq f_{\rm max}$ and 
$Q_{\rm min} \leq Q \leq Q_{\rm max}$ is estimated by integrating
the volume element $\sqrt{\det g}$ 
over that region of the template space. 
We find
\begin{eqnarray}
N_{\rm f} &=& {\displaystyle \frac {1}{2}} (ds^2_{\rm max})^{-1} 
\,\mathrm{ln} ({f_{\rm max}/f_{\rm min}})
\nonumber \\ && \quad \times 
\int_{Q_{\rm min}}^{Q_{\rm max}} dQ
\,{\displaystyle \frac 
{Q\,(16\,Q^{6} + 32\,Q^{4} + 10\,Q^{2} + 1)^{1/2}}
{(2\,Q^{2} + 1)^{3/2}(4\,Q^{2} + 1)^{1/2}}} \,.
\label{eq:Nf}
\end{eqnarray}
The $Q_{\rm max}$-dependence of $N_{\rm f}$ is shown in Fig.~\ref{fig:Qdep}.
For large $Q_{\rm max}$, it can be derived analytically as~\cite{crei}
\begin{eqnarray}
\left. N_{\rm f}\right|_{{\rm large-}Q} 
= {\displaystyle \frac {\sqrt{2}}{4}} (ds^2_{\rm max})^{-1} 
\,\mathrm{ln} ({f_{\rm max}/f_{\rm min}})\,Q_{\rm max} \,.
\label{eq:Lq}
\end{eqnarray}

\begin{figure}[ht]
\center
\epsfxsize=8cm
\hspace{-1cm}\epsfbox{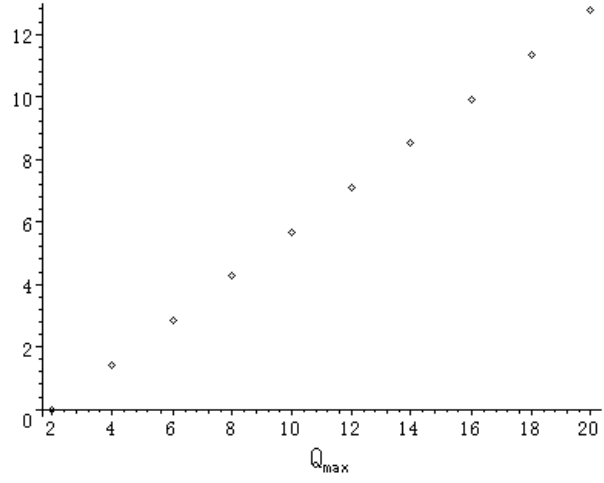}
\caption{The $Q_{\rm max}$-dependence of the number of required filters
[$2\,ds^2_{\rm max} N_{\rm f}/\log({f_{\rm max}/f_{\rm min}})$].
Here we chose $Q_{\rm min}=2$. }
\label{fig:Qdep}
\end{figure}

{}From Fig.~\ref{fig:Qdep}, we see that
the large $Q_{\rm max}$ approximation is good even for smaller $Q_{\rm max}$
as far as the number of required filters is concerned. 
In Table~\ref{table:temp}, we show the number of filters
in the case $Q_{\rm min}=2$, $f_{\rm min}=100{\rm Hz}$, 
and $ds^2_{\rm max}=0.03$ for several values of 
$f_{\rm max}$ and $Q_{\rm max}$. 
We also note that the number of required filters depends 
on the choice of a tiling method in practice.
\begin{table}[ht]
\begin{center}
\renewcommand{\arraystretch}{1.8}
  \begin{tabular}{|c||c|c|c|}
\hline
$f_{\rm max}$(Hz) & $Q_{\rm max}=10$
& $Q_{\rm max}=10^2$ & $Q_{\rm max}=10^3$ \\
    \hline 
$10^3$
& $218$ & $2,661$ & $27,084$ 
 \\ \hline
$10^4$
& $436$ & $5,321$ & $54,167$
 \\
\hline
  \end{tabular}
 \end{center}
 \caption{The estimation of the number of filters with the maximum SNR loss
of $3\,\%$}
\label{table:temp}
\end{table}

Next, let us perform the coordinate transformation by which 
the cosine-wave template metric~(\ref{eq:ds2c}) is transformed to 
a diagonal, conformally flat metric. 

We start from the coordinate transformation that
removes the frequency dependence in the metric, 
\begin{eqnarray}
dF = d\ln f_c \,, 
\end{eqnarray}
which gives
\begin{eqnarray}
ds^2 &=&  g_{FF} \,dF^{2} 
+ g_{QQ} \,dQ^{2}
+2\,g_{FQ} \,dF\,dQ 
\nonumber\\
 &=& {\displaystyle \frac {1}{8}} \,
{\displaystyle \frac {16\,Q^4+6\,Q^{2} + 1}{(2\,Q^{2} + 1)}} \,dF^{2} 
+ {\displaystyle \frac {1}{8}} \,
{\displaystyle \frac {64\,Q^8+128\,Q^{6} + 28\,Q^{4} + 1}
{Q^2\,(4\,Q^{2} + 1)^{2}(2\,Q^{2} + 1)^{2}}} \,dQ^{2}
\nonumber \\ &&
\mbox{}  - 
{\displaystyle \frac {1}{4}} \,
{\displaystyle \frac {8\,Q^{4} +2\,Q^{2}+ 1}
{Q\,(4\,Q^{2} + 1)(2\,Q^{2} + 1)}} \,dF\,dQ \,.
\label{eq:ds2nn}
\end{eqnarray}
The transformation that removes the off-diagonal element
is found by setting $F=X-u(Q)$ and
requiring
\begin{eqnarray}
g_{FF}\,u'(Q)-g_{FQ}=0\,.
\label{eq:udef}
\end{eqnarray}
We find
\begin{eqnarray}
u={3\sqrt{7}\over14}
\left[
\arctan\left({3+16Q^2\over\sqrt{7}}\right)-{\pi\over2}
\right]
-{1\over4}\ln{16Q^4(16Q^4+6Q^2+1)\over(4Q^2+1)^4}\,,
\label{eq:usol}
\end{eqnarray}
which gives
\begin{eqnarray}
ds^2
=g_{FF}\,dX^2
+\left(g_{FF}\,u'{}^2-2g_{FQ}\,u'+g_{QQ}\right)dQ^2
=g_{FF}\left(dX^2+{g_{FF}\,g_{QQ}-g_{FQ}^2\over g_{FF}^2}\right)dQ^2\,.
\end{eqnarray}
Then we can perform a further coordinate transformation to make
the metric conformally flat. Namely, by the transformation
$Q\to Y$ defined by
\begin{eqnarray}
Y=\int_{Q}^\infty dQ'{\sqrt{\det g(Q')\over g_{FF}(Q')}}
=\int_{Q^2}^\infty dx\,
{2\sqrt{64x^4+144x^3+72x^2+14x+1}\over
(64x^3+40x^2+10x+1)\sqrt{1+2x}}
\quad(Q>0),
\label{eq:Ydef}
\end{eqnarray}
we obtain
\begin{eqnarray}
ds^2=\Omega(Y)\left(dX^2+dY^2\right)\,,
\label{eq:ds2con}
\end{eqnarray}
where the conformal factor is given by
\begin{eqnarray}
\Omega(Y)=g_{FF}\bigl(Q(Y)\bigr).
\end{eqnarray}
Here $Q$ is now a function of $Y$ determined by inverting
Eq.~(\ref{eq:Ydef}).

Although the above coordinate transformation involves complicated
functions that may not be expressed in terms of elementary functions,
we find it is sufficient to use their large $Q$ expansion forms 
for $Q\geq2$.
Up to $O(1/Q^8)$ inclusive, we have
\begin{eqnarray}
X&=&F+u=F+ {1 \over 16}{1 \over Q^2}
-{3 \over 256}{1 \over Q^4}+{13 \over 3072}{1 \over Q^6}
-{43 \over 32768}{1 \over Q^8}+{109 \over 327680}{1 \over Q^{10}} \,, \\ 
Y &=& {\sqrt{2} \over 4}
\left({1 \over Q}
+{1 \over 12}{1 \over Q^3}-{73 \over 640}{1 \over Q^5}
+{21 \over 256}{1 \over Q^7}-{6127 \over 98304}{1 \over Q^9}
\right) \,.
\end{eqnarray}
To the same accuracy, the inverse transformation becomes
\begin{eqnarray}
F &=& X-{1 \over 2}{Y^2}
+{17 \over 12}{Y^4}-{586 \over 45}{Y^6}
+{47587 \over 360}{Y^8}-{3153013 \over 2025}{Y^{10}} \,, 
\\ 
Q &=& {\sqrt{2} \over 4}
\left(
-{4401959 \over 5400}{Y^7}
+{9892 \over 135}{Y^5}-{737 \over 90}{Y^3}
+{2 \over 3}Y+{1 \over Y}\right) \,.
\end{eqnarray}
The conformal factor $\Omega(Y)$ is given by
\begin{eqnarray}
\Omega(Y) &=& -{749639 \over 5400}{Y^6}
+{1255 \over 108}{Y^4}-{119 \over 120}{Y^2}
+{1 \over 24}+{1 \over 8}{1 \over Y^2} \,.
\end{eqnarray}

The structure of this metric is the same as the one in the general 
case discussed in Section III. 
The effective tiling method describe in Section III can thus be used 
to set the template space using this metric. 

%%%%%%%%%%%%%%%%%%%%%%%%%%%%%%%%%%%
\subsection{Sine wave} \label{app:sin}
%%%%%%%%%%%%%%%%%%%%%%%%%%%%%%%%%%%

Next, we consider the damped sine wave 
$e^{ - \frac {\pi \,f_c\,t}{Q}}\,\sin(2\,\pi \,f_c\,t)$. 
The calculation is done as same as 
the cosine wave case. 
First, we obtain 
\begin{eqnarray}
ds^2 =  {\displaystyle \frac {1}{8}} \,
(8\,Q^2+3) \,dF^{2} 
+ {\displaystyle \frac {1}{8}} \,
{\displaystyle \frac {16\,Q^4+3}
{Q^2\,(4\,Q^{2} + 1)^{2}}} \,dQ^{2}
-{\displaystyle \frac {1}{4}} \,
{\displaystyle \frac {4\,Q^{4} +3}
{Q\,(4\,Q^{2} + 1)}} \,dF\,dQ \,.
\end{eqnarray} 
For the above metric, 
the $1/Q$-expansion is performed and we obtain
\begin{eqnarray}
ds^2 &=&  g_{FF} \,dF^{2} 
+ g_{QQ} \,dQ^{2}
+2\,g_{FQ} \,dF\,dQ \,, 
\label{eq:ds2nnss} \\ 
g_{FF} &=& Q^2+{3 \over 8} \,,
\nonumber \\ 
g_{QQ} &=& {1 \over 8}{1 \over Q^2}
-{1 \over 16}{1 \over Q^4}+{3 \over 64}{1 \over Q^6}
-{5 \over 256}{1 \over Q^8}+{7 \over 1024}{1 \over Q^{10}}
\,,
\nonumber \\ 
g_{FQ} &=& -{1 \over 8}{1 \over Q}
-{1 \over 16}{1 \over Q^3}+{1 \over 64}{1 \over Q^5}
-{1 \over 256}{1 \over Q^7}+{1 \over 1024}{1 \over Q^9}
\,. \nonumber
\end{eqnarray} 
The expansion is considered up to $O(1/Q^8)$ beyond the leading term. 

Next, we consider the following coordinate transformation, 
\begin{eqnarray}
X &=& F 
+{1 \over 16}{1 \over Q^2}+{1 \over 256}{1 \over Q^4}
-{11 \over 3072}{1 \over Q^6}
+{49 \over 32768}{1 \over Q^8}-{179 \over 327680}{1 \over Q^{10}} \,, \\ 
Y &=& {\sqrt{2} \over 4}
\left({1 \over Q}-{1 \over 6}{1 \over Q^3}
+{27 \over 640}{1 \over Q^5}
-{43 \over 3584}{1 \over Q^7}+{1067 \over 294912}{1 \over Q^9}
\right) \,.
\end{eqnarray}
The inverse transformation of this is given by 
\begin{eqnarray}
F &=& X-{1 \over 2}Y^2-{19 \over 12}Y^4
-{136 \over 45}Y^6-{15263 \over 2520}Y^8
-{181651 \over 14175}Y^{10} \,, 
\\ 
Q &=& {\sqrt{2} \over 4}
\left(
-{71033 \over 37800}Y^7-{1073 \over 945}Y^5
-{77 \over 90}Y^3-{4 \over 3}Y+{1 \over Y}\right) \,.
\end{eqnarray}

Finally, we obtain the conformally flat metric as 
\begin{eqnarray}
ds^2 &=& \Omega(Y) \left(dX^2 + dY^2 \right) \,,
\\
\Omega(Y) &=& {1 \over 5400}Y^6+{1 \over 756}Y^4+{1 \over 120}Y^2
+{1 \over 24}+{1 \over 8}{1 \over Y^2} \,.
\end{eqnarray}
The structure of this metric is the same as the one in the 
the general case of Section III and the case 
of cosine wave.
Especially, the leading term with respect to $1/Q$ are the same
as in the case of cosine wave. 
Therefore, the effective tiling method described in Section III is
also applicable to this case. 

%%%%%%%%%%%%%%%%%%%%%%%%%%%%%%%%%%%
%%%%%%%%%%%%%%%%%%%%%%%%%%%%%%%%%%%
\section{The number of templates when the mass is known}\label{sec:IR}
%%%%%%%%%%%%%%%%%%%%%%%%%%%%%%%%%%%
%%%%%%%%%%%%%%%%%%%%%%%%%%%%%%%%%%%

In this appendix, we consider the case when a compact star 
with mass $1\sim 10^2M_\odot$
is inspiraling into a super massive black hole with
mass $10^6\sim 10^8M_\odot$, and after the 
final plunge, the ringing wave is excited. 
In such cases, it is expected that 
masses of the super massive black hole and the compact star
is determined accurately during the inspiral phase. 
We do not know exactly the mass of the final black hole 
after the plunge of the compact star because 
we do not know how much energy are radiated as the gravitational 
waves and how much mass loss will be occurred in the case of neutron star. 
In this case, however, we can assume the mass of 
the final black hole within the accuracy $10^{-8}\sim 10^{-4}$
since the mass ratio is very large. 
In this case, we can perform an analysis to search 
for the ringdown wave using these mass parameters. Here we investigate
how much templates we need in this situation. 

When we know the mass $M$ of a black hole, we only need to investigate
the spin of the black hole. 
{}From Eqs. (\ref{eq:fcMa}) and (\ref{eq:QMa}), the relation between $f_c$ and $Q$ becomes 
\begin{eqnarray}
f_c &\simeq& 320\,
\left[1-0.63\left({Q \over 2}\right)^{-2/3}\right]
\left({M \over M_{\odot}}\right)^{-1} 
\label{eq:relfcQ}\,,
\end{eqnarray}
where we normalize the frequency by $100$Hz. 
Now, we consider the white noise case. We only consider the cosine wave case. 
These assumptions are sufficient for the purpose here. 
{}From Eq.~(\ref{eq:relfcQ}), 
\begin{eqnarray}
{\partial f_c \over \partial Q} 
= 210\left({M \over M_{\odot}}\right)^{-1} Q^{-5/3} \,.
\end{eqnarray}
The metric (\ref{eq:ds2c}) in the large $Q$ limit is derived as 
\begin{eqnarray}
ds^2 &\sim& \left(0.45\,Q^{-4/3}-0.17\,Q^{-8/3}+0.13\,Q^{-2}\right)dQ^2 
\nonumber \\ 
&\sim& 0.45\,Q^{-4/3} dQ^2 \,.
\end{eqnarray}
When the maximum loss of the signal to noise ratio 
$ds^2_{\rm max}$ is given, 
the number of templates needed to achieve this efficiency is estimated as 
$N\simeq (Q^{1/3}_{\rm max}-Q^{1/3}_{\rm min})/ds_{\rm max}$. 
If we consider $Q_{\rm min}=2$ and $Q_{\rm max}=100$, we only need 20 templates. 
Thus, the prior knowledge of the black hole mass will make
the matched filtering analysis substantially easy.

\end{appendix}

%%%%%%%%%%%%%%%%%%%%%%%%%%%%%%%%%%%%%%%%
%%%%%%%%%%%%%%%%%%%%%%%%%%%%%%%%%%%%%%%%

\end{document}